\documentclass[fleqn]{elsart}

\usepackage{amsmath,amssymb,amsfonts}
\usepackage{bm}

\newcommand{\be}{\begin{equation}}
\newcommand{\ee}{\end{equation}}
\newcommand{\ba}{\begin{eqnarray}}
\newcommand{\ea}{\end{eqnarray}}
\def\bs{\begin{subequations}}
\def\es{\end{subequations}}
\def\a{\alpha}
\def\b{\beta}
\def\g{\gamma}
\def\e{\epsilon}
\def\t{\theta}
\def\s{\sigma}
\def\cE{{\cal E}}

\def\p{\partial}
\newcommand{\Eq}[1]{(\ref{#1})}

\begin{document}

\begin{frontmatter}

\rightline{\small IGC-08/6-5 \hfill}
\vspace{1cm}

\title{Superconducting loop quantum gravity and the cosmological constant}

\author{Stephon H.S. Alexander},
\ead{stephonalexander@mac.com}
\address{Institute for Gravitation and the Cosmos, Department of Physics,\\ The Pennsylvania State University,
104 Davey Lab, University Park, PA 16802}
\address{Department of Physics, Haverford College, 370 Lancaster Avenue, Haverford, PA 19041}
\author{Gianluca Calcagni},
\ead{gianluca@gravity.psu.edu}
\address{Institute for Gravitation and the Cosmos, Department of Physics,\\ The Pennsylvania State University,
104 Davey Lab, University Park, PA 16802}
\date{June 26, 2008}
\begin{abstract}
We argue that the cosmological constant is exponentially suppressed in a candidate ground state of loop quantum gravity as a nonperturbative effect of a holographic Fermi-liquid theory living on a two-dimensional spacetime. Ashtekar connection components, corresponding to degenerate gravitational configurations breaking large gauge invariance and CP symmetry, behave as composite fermions that condense as in Bardeen--Cooper--Schrieffer theory of superconductivity. Cooper pairs admit a description as wormholes on a de Sitter boundary.
\end{abstract}


\end{frontmatter}


If the observed dark energy is associated with all the contributions from quantum fields to the cosmological constant $\Lambda$, we have to explain why these are suppressed so as to render the vacuum energy $\Lambda \sim 10^{-120}$ (in reduced Planck units). The dark energy problem is further obscured when the issue of general covariance arises. For example, a zero occupation number vacuum state is not invariant under general coordinate transformations \cite{unr76}. One approach is to find a mechanism wherein the fundamental degrees of freedom of quantum gravity dynamically regulate the cosmological constant at the level of general covariance. Once a spacetime background is specified, we should be able to identify the root of the ambiguity that we currently face in the evaluation of dark energy.    
 
How does one truly deal with the cosmological constant problem in a background independent manner unless one includes matter fields? Here we entertain the possibility that the vacuum energy evaluated on a degenerate spacetime is responsible for a dynamical suppression of $\Lambda$. This sector violates parity and is described, at the quantum level, by a model with a four-fermion interaction which reproduces that of Cooper pairing in Bardeen--Cooper--Schrieffer (BCS) theory \cite{BCS}. This correspondence is actually more general and it can be shown that gravity allows for a dual description in terms of a nonlocally interacting Fermi liquid; which is done in a companion paper, where the reader will find all the ingredients of this picture reviewed or developed in greater detail \cite{SC}.

To facilitate the study of the cosmological constant we employ the Ashtekar formalism \cite{ash86} of classical general relativity, which is described by a complex connection field $A\equiv A_\a dx^\a\equiv A_\a^i \tau_i dx^\a$ and a real triad $E$ obeying a canonical equal-time Poisson algebra $\{A_\a^i({\bf x}),\,E_j^\b({\bf y})\}=i\delta_\a^\b\delta_j^i\delta({\bf x},{\bf y})$, where Greek indices $\a,\b,\dots$ denote spatial components, Latin letters $i,j,\dots$ label directions in the internal gauge space, and $\tau_i$ are generators of the $su(2)$ gauge algebra. Introducing the gauge field strength $F_{\a\b}^k\equiv \p_\a A_\b^k-\p_\b A_\a^k+\e_{ij}\phantom{}\phantom{}^{k}A_\a^iA_\b^j$ and the magnetic field $B^{\a i}\equiv \e^{\a\b\g}F_{\b\g}^i/2$ ($\epsilon_{\a\b\g}$ is the Levi--Civita symbol), the scalar, Gauss, and vector gravitational constraints in the presence of a cosmological constant are
\ba
{\cal H} &=& \e_{ijk} E^i\cdot E^j\times \left(B^k+\frac{\Lambda}3E^k\right)=0\,,\nonumber\\
{\cal G}_i &=& D_\a E^\a_i=0\,,\nonumber\\
{\cal V}_\a&=& (E_i\times B^i)_\a=0\nonumber,
\ea
where $({\bf a}\times {\bf b})_\a= \e_{\a\b\g}a^\b b^\g$ and $D_\a$ is the covariant derivative. The Gauss constraint guarantees invariance under gauge transformations homotopic to the identity; the total Hamiltonian can be made invariant also under large gauge transformations by shifting the momentum $E$ by the axial magnetic field \cite{ABJ,mon01}.

A candidate background-independent quantum theory of gravity at small scales is loop quantum gravity (LQG) \cite{rov04}, which will provide the necessary interpretational framework. The triad becomes the operator $\hat E^\a_i= -\delta/\delta A_\a^i$, while $\hat A_\a^i$ is multiplicative in the naive connection representation. The quantum constraints on a kinematical state $\Psi(A)$ read
\ba
\hat{\cal H}\Psi(A) &=& \e_{ijl}\e_{\a\b\g} \frac{\delta}{\delta A_{\a i}}\frac{\delta}{\delta A_{\b j}} \hat{\cal S}^{\g l}\Psi(A)=0,\label{qh}\\
\hat {\cal G}_i\Psi(A) &=& -D_\a \frac{\delta}{\delta A_{\a i}}\Psi(A)=0,\label{qg}\\
\hat{\cal V}_\a\Psi(A)&=& -\frac{\delta}{\delta A_\b^i}F_{\a\b}^i\Psi(A)=0,\label{qv}
\ea
where ${\cal S}^{\g l}\equiv B^{\g l}+(2k/\pi)E^{\g l}$ and
\be\label{k}
k\equiv\frac{6\pi}{\Lambda}
\ee
in Lorentzian spacetime, and we adopted a factor ordering with the triads on the left \cite{ash86,BGP3}, which makes the quantum constraint algebra consistent. A solution to all constraints (in their smeared form) is the Chern--Simons state \cite{kod90} (see \cite{smo02,GP} for reviews)
\be\label{cs}
\Psi_{\rm CS}={\cal N} \exp\left(\frac{k}{4\pi}\int_{S^3} Y_{\rm CS}\right)\,,
\ee
where ${\cal N}$ is a normalization constant and $Y_{\rm CS}$ is the Chern--Simons form 
\be
Y_{\rm CS}\equiv \frac12{\rm tr}\left(A \wedge dA+\frac23 A \wedge A \wedge A\right),
\ee
on the 3-sphere. Solution of the Hamiltonian constraint is guaranteed by the property
\be
\frac{\delta}{\delta A_{\g k}}\int_{S^3} Y_{\rm CS}=2B^{\g k}\,.\nonumber
\ee
Non-Abelian gauge theories like QCD made invariant under large gauge transformations admit instantonic degrees of freedom that have phenomenological consequences \cite{CDG,JR}. Likewise, the Chern--Simons ground state encodes degrees of freedom connecting families of spacetimes. In Euclidean gravity ($k\equiv i\t/\pi\to i6\pi/\Lambda$), different sectors are connected by unitary large gauge transformations which shift the topological phase $\t$ by integer values. Under a large gauge transformation characterized by winding number $n$, $\int Y_{\rm CS}\to \int Y_{\rm CS}+4\pi^2n$ \cite{wei}, and the Chern--Simons state transforms as $\Psi_{\rm CS}(A)\to e^{i n\t}\Psi_{\rm CS}(A)$ \cite{ABJ}.

The inclusion of matter perturbations reproduce standard quantum field theory on de Sitter background \cite{SS}, while linearizing the quantum theory one recovers long-wavelength gravitons on de Sitter \cite{smo02}. In this sense the Chern-Simons state is a genuine ground state of the theory. When the connection is real, the Chern--Simons state closely resembles a topological invariant of knot theory \cite{wit89}. This identification opens up a possibility of describing quantum gravitational dynamics with the mathematics developed in knot theory \cite{BM}. Also, it is generally believed that matter excitations might be realized as particular states in the spin network space, where braid configurations (corresponding to standard model generations) are expected to live \cite{BMS}.

Despite these and other beautiful properties, several issues have been raised against the Chern--Simons state, including the problem of normalizability ($\Psi_{\rm CS}$ is not normalizable), reality ($A$ is self-dual), and representation (we still lack a suitable measure in configuration space allowing one to resort to the LQG holonomy representation). Nevertheless, the Chern--Simons state has passed several independent consistency checks \cite{GP}, and many of the above objections have been addressed at least partly in recent investigations \cite{soo01,ran2}.

Whenever a cosmological constant component is required by observations at some point during the evolution of the universe, agreement with physical observables lead to the phenomenological hypothesis that the vacuum component is actually dynamical; its behaviour can be reproduced by the matter (typically, scalar) field characteristic of quintessence and inflationary models. In this respect what we are going to do, promoting $\Lambda$ to an evolving functional $\Lambda(A)$, is not uncommon. Nonetheless, the justification and consequences of this step change perspective under the lens of loop quantum gravity. Looking at Eq.~\Eq{k}, one can see that a dynamical cosmological constant corresponds to a deformation of the topological sector of the quantum theory:
\be\label{ta}
k\to k(A)\,.
\ee
In QCD the partition function can be extended to a larger $U(1)$ symmetry, namely the Peccei--Quinn invariance under a rotation by the $\theta$ angle \cite{PQ1}. Instanton effects can spontaneously break the $U(1)$ symmetry resulting in a light particle called axion. Likewise, a large gauge transformation in Chern--Simons wavefunction is regarded as a $U(1)$ rotation, so by deforming $\theta$ we break this symmetry. Classically, unless $k$ is invariant under small gauge transformations, the only sectors compatible with Eq.~\Eq{ta} and the Gauss constraint are degenerate ($\det E=0$). We exclude the most degenerate case ${\rm rk} E=0$, as we want to preserve at least part of the canonical algebra. This leaves the cases ${\rm rk} E=1,2$.

The deformed quantum scalar constraint is defined with $\Lambda(A)$ to the left of triad operators; however, if the triad operator is degenerate the scalar constraint plays no dynamical role. Therefore it is consistent to assume the same attitude as in the discussion of the Chern--Simons state, and require that the deformed state $\Psi_*$ annihilates the deformed reduced constraint ${\cal S}_*^{\a i}$. The latter requires the addition of a counterterm,
\be
(\hat\Theta^{\a i}+\hat{\cal S}_*^{\a i})\Psi_*=\hat \Theta^{\a i}+\frac12\int_{S^3} Y_{\rm CS}\frac{\delta \ln \Lambda(A)}{\delta A_{\a i}}=0\,,\label{th0}
\ee
which breaks large gauge invariance. (Nonlocal effects, expected from this symmetry breaking, will be discussed later.) The equation of motion for the gauge field becomes
\ba
\dot A_\a^i &=& i\e^i_{\phantom{i}jk}E^{\b j}\left[F_{\a\b}^k+ \frac{\Lambda}2\,\e_{\a\b\g} E^{\g k}-\e_{\a\b\g}\int_{S^3} Y_{\rm CS}\frac{\delta \ln \Lambda(A)}{\delta A_{\g k}}\right]\,.\label{eom2}
\ea
We now make a crucial connection which was demonstrated by Jacobson for classical gravity \cite{jac96}. 
One of the advantages of the formulation of classical gravity via Ashtekar variables is the possibility to have a well-defined causal structure \cite{mat95} even when the background metric is singular, as inside a black hole, at big-bang or big-crunch events, or in processes where topology changes. In particular, the degenerate sector with ${\rm rk} E=1$ describes a two-dimensional spacetime whose future is a tipless wedge. The gravielectric line can be chosen to lie on the $z$ direction; after some other gauge fixing, Jacobson's sector read
\bs\label{des}\ba
&& A_z^i=0\,,\quad A_a^3=A_a^3(x^a)\,,\quad A_a^{i\neq 3}=A_a^{i\neq 3}(t,z)\,,\\
&& E^z_{i\neq 3}=0\,,\qquad E^a_i=0\,,\qquad E^z_3=1\,.
\ea\es
The classical equation of motion for the transverse-transverse components of the connection is $\dot A_a^i= -i\epsilon^{i3}_{\phantom{i3}j} \p_z A_a^j$, which can be written as the $(1+1)$-dimensional Dirac equation
\be\label{dieq}
\g^0\dot\psi+\g^z\p_z\psi=0\,,
\ee
where $\g^\mu$ are Dirac matrices and
\be\label{psia}
\psi\equiv\begin{pmatrix} iA^1_1 \\ A^1_2 \\ A^2_1 \\ iA^2_2\end{pmatrix}\,.
\ee
One can take differently oriented gravitational lines and patch them together at their boundaries. The emerging classical picture is that of two-dimensional worldsheets, similar to fermionic string worldsheets, that communicate at the edges, where the gravitational field may be nondegenerate. Geometry amounts to a collection of gravitational lines on which the triad $E^\a_i$ is a constant vector and vanishes elsewhere. However, one lacks a model for this interaction, as well as a physical interpretation of the worldsheet network and fermionic degrees of freedom. Both naturally emerge at the quantum level when the deformed constraints act upon the Chern--Simons state. 

Let us go back to the effective Hamiltonian of quantum gravity with the nonperturbative counterterm. At this point we show that an exponentially suppressed cosmological constant provides a nonperturbative solution of the fermion manybody system. Taking $\Lambda(\psi)=\Lambda_0 \exp (\xi^{T}\psi)$,
where $\xi^{T}=(\xi_1,\xi_2,\xi_3,\xi_4)$ is an arbitrary field and $\Lambda_0=O(1)$ is an integration constant, in Jacobson's sector Eq.~\Eq{eom2} becomes \cite{SC} $\g^0\dot\psi+\g^z\p_z\psi+i\widetilde m\g^x\xi=0$,
where $\widetilde m\equiv -i\psi^T\g^x\g^z\p_z\psi$. This effective mass is nonperturbatively generated by the anisotropic cross-interaction of the connection components. Imposing the condition $\xi=-\g^x\psi$, the field $\psi$ would obey a Dirac equation 
\be\label{eom4}
\g^0\dot\psi+\g^z\p_z\psi+im\psi=0\,,
\ee
with mass $m=2\widetilde m$. Classically the mass would vanish, as $\xi^T\psi=0$. However, after quantizing the field its components become Grassmann variables and the symmetry-breaking mechanism comes into effect.\footnote{This entails
the identification of a bosonic field (connection components) with a fermionic object (spinor components). The issue of spin statistics will be pursued in another paper.} In order to get quantities with well-defined Lorentz structure, $\psi^T$ should actually be related to $\bar\psi\equiv\psi^\dagger\g^0$. This is realized if $\psi=\psi_c\equiv -i\g^y\psi^*$, i.e., if $\psi$ is equal to its charge conjugate (Majorana fermion). Then the connection components must satisfy the conditions $A_1^2=A_2^{1*}$, $A_2^2=A_1^{1*}$, and $\Lambda$ is
\be\label{lam}
\Lambda=\Lambda_0 \exp(-\bar\psi\g^5\g^z\psi)\,,
\ee
where $\g^5\equiv i\g^0\g^x\g^y\g^z$. The cosmological constant encodes the imprint of an axial vector current $j^{5\a}$, associated with a chiral transformation of the fermion $\psi$ and not conserved in the presence of the effective mass
\be\label{mass2}
m= -2i\bar\psi\g^5\p_z\psi\,.
\ee
The deformation process affects the topological sector and the CP symmetry of the theory. Equation \Eq{lam} can address the related smallness problem in terms of a condensate with vacuum expectation value (with respect to $\Psi_*$)  $\langle j^{5z}\rangle\sim O(10^2)$. In the perturbative regime (small values of the connection, $|\langle j^{5z}\rangle|\ll 1$), $\langle\Lambda\rangle\approx\Lambda_0(1-\langle j^{5z}\rangle)=O(1)$. In the nonperturbative regime, the effective mass grows important and the cosmological constant, supposing $\langle j^{5z}\rangle$ to be positive definite for large connection values, becomes exponentially small. Thus the smallness of the cosmological constant is regarded as a large-scale nonperturbative quantum mechanism similar to quark confinement.

We propose Eqs.~\Eq{eom4} and \Eq{mass2} as the starting point of a physical reinterpretation of quantum gravity. One can canonically quantize $\psi$ and expand it in discrete one-dimensional momentum space, $\psi = \sum_{k,\s} e^{ikz} (2\cE_k)^{-1/2} (c_{k\s}u_{k\s}+c_{-k\s}^\dagger v_{k\s})$, where $\s=\pm$ is the spin, $\cE_k$ is the energy, $c$ and $c^\dagger$ are annihilation and creation operators obeying a fermionic algebra $\{c_{k\s},c_{k'\s'}^\dagger\}=\delta_{kk'}\delta_{\s\s'}$, and $u$ and $v$ are spinorial functions. Due to Eq.~\Eq{mass2}, in the Hamiltonian there appear fermionic nonlocal correlations of the form
\be
\sum_{\s,k,k'}V_{kk'}c^\dagger_{k\s} c^\dagger_{k'\s'} c_{k\s} c_{k'\s'}\,,
\ee
where $V_{kk'}$ is a function of the momenta and is determined by the effective mass \Eq{mass2}. Spin models of this form, generically called Fermi-liquid theories, are employed in condensed matter physics to describe fermionic systems with many-body interactions. In particular, in mean-field BCS theory ($V_{kk'}=g=$const.) fermions with opposite spin can nonlocally interact in $N$ pairs at a given energy level, lower than the Fermi energy of the free-fermion sea, leading to a medium with superconducting properties. This is regarded as the true vacuum of the theory, $|{\rm BCS}\rangle = \exp\sum_{k}(v_k/u_k) c_{+}^\dagger c_{-}^\dagger|0\rangle_N$, where $|0\rangle_N$ is the $N$-pairs vacuum. It is separated from the perturbative vacuum by a mass gap $\Delta\sim e^{-1/g}$ (in the weakly-coupled regime). One can check that the gravitational analogues of the BCS wavefunction and mass gap are the Chern--Simons state and the cosmological constant.

The relationship between BCS theory and degenerate LQG can be better formalized by recognizing the former as a deformed CFT, namely, an $SL(2,\mathbb{R})_{k=2}$ WZW model \cite{WZ,W} at critical level with a vertex operator breaking conformal invariance \cite{sie99}. Hamiltonian BCS eigenstates are in correspondence with deformed conformal blocks, while the creation-annihilation pair operators are in one-to-one correspondence with the charges $Q^{0,\pm}\sim\oint dz J^{0,\pm}$ associated with the Sugawara currents of the WZW model in Wakimoto representation.

On the other hand, this is the same structure one encounters in quantum spin networks (e.g., \cite{smo02}), where edges are two-dimensional tubular surfaces carrying an irreducible representation of the $q$-deformed algebra $su(2)_k$, vertices are promoted to punctured two-spheres, and punctures are described by WZW conformal blocks. Therefore, BCS levels are mapped to edges of quantum spin networks, interacting at nodes via a BCS coupling. On $N$ edges there lives a copy of the identity element represented by a WZW screening charge corresponding to a Cooper pair. The interaction among pairs is given by an operator breaking conformal invariance and evaluated along a loop encircling all the pairs. 

The BCS interaction is the cornerstone of the construction of three-dimensional geometry from quantum spin networks. Classically, it describes scattering of Jacobson electring lines at their endpoints; these worldsheets, patched together at their edges, span the three-dimensional space, giving rise to the geometric sector which was lost in the free-field picture. 
Before quantizing the theory, we fixed the gravitational configuration to be a particular degenerate sector, namely Jacobson's. This way of proceeding is similar to that of loop quantum cosmology and minisuperspace models, where a reduction of the symplectic structure is performed at classical level. However, it does not result in a loss of generality in the present framework, as we have just argued.

The screening charges at a given node have an intuitive picture as the sites activated in an area measurement, i.e., when a classical area intersects the spin network. In condensed matter physics, the Fermi sea is a ground state of uncorrelated electron pairs whose Fermi energy is higher that the BCS pair-correlated state. In quantum gravity, pair correlation can be regarded as a process of quantum decoherence. An abstract spin network is the gravity counterpart of an unexcited Fermi sea. As soon as an area measurement is performed on the state, $N$ edges of a given node are activated and the system relaxes to a lower-energy vacuum corresponding to the selection of one of the area eigenstates in a wave superposition. In a sense, physical areas are Fermi-liquid surfaces and measuring geometry equals to counting Cooper pairs.

Given these forms of evidence, we conjecture that \emph{Loop quantum gravity with a cosmological constant is dual to a two-dimensional Fermi liquid living on an embedded spin network}. The classical spacetime corresponding to the Chern--Simons state is de Sitter, whose entropy depends on the horizon area. This result is explained in terms of the microscopic degrees of freedom at the boundary, which are nothing but Cooper pairs or, in the language of gravity, oppositely charged pairs of singularities (wormholes) opening and closing on the boundary. This type of classical solution, which reproduces Jacobson's sector, was found in \cite{Kam91}, where the magnetic field $B^i_\a$ on the horizon is sourced by a monopole-antimonopole configuration. Ergo, another way to state our conjecture is: \emph{Loop quantum gravity with a cosmological constant is dual to a two-dimensional Fermi liquid living on the time-space boundary of a de Sitter background}.

Quantum gravity in a four-dimensional causal degenerate sector with a deformed topological structure mimics the behaviour of lower-dimensional Fermi liquids and, in the simplest case, is actually equivalent to a BCS theory of superconductivity. It is known that the nonlocal nature of LQG prevents the formation of what are classically regarded as singularities \cite{Bo06a,Bo06b,Ash08}; roughly speaking, quanta of geometry cannot be compressed too densely and they determine the onset of a repulsive force at Planck scale \cite{Ash08}. Here we have seen that the agent behind this mechanism is literally Pauli exclusion principle. The details of the possible multi-fermion interactions in gravitational degenerate sectors will require further attention and the wisdom of condensed matter models. Further generalizations to non-BCS models, including non-mean-field Fermi liquids (corresponding to WZW noncritical quantum algebras), bosonic condensates (related to the emergence of matter degrees of freedom), and BEC-BCS crossover (related to decoherence and the problem of measurement) will shed some light into the fermionic nature of quantum gravity and its weak/strong-coupling relations.


\ack
We thank A. Ashtekar, M. Bojowald, A. Corichi, L. Freidel, R. Gambini, E.E. Ita, T. Jacobson, J.K. Jain, F.R. Klinkhamer, S. Mercuri, J. Pullin, A. Randono, G. Sierra, J. Sofo, and especially J.D. Bjorken and L. Smolin for useful discussions. S.H. is supported by NSF CAREER Award and G.C. by NSF grant PHY-0653127 and partly by Perimeter Institute for Theoretical Physics.


\end{document}